\documentclass[final,5p,times,twocolumn]{elsarticle}

\usepackage{amssymb}
\newcommand{\pnpnpippim}{$np \to np\pi^+\pi^-$}
\newcommand{\ppnnpippip}{$pp \to nn\pi^+\pi^+$}
\newcommand{\pim}{$\pi^-$}
\newcommand{\pip}{$\pi^+$}
\newcommand{\piz}{$\pi^0$}
\newcommand{\Del}{$\Delta$}
\newcommand{\nstar}{$N(1440)$}
\journal{Physics Letters B}

\begin{document}

\begin{frontmatter}



\title{Study of the quasi-free $np \to np \pi^+\pi^-$ reaction with a deuterium beam at 1.25 GeV/nucleon}


\address[a]{Istituto Nazionale di Fisica Nucleare - Laboratori Nazionali del Sud, 95125~Catania, Italy}
\address[b]{LIP-Laborat\'{o}rio de Instrumenta\c{c}\~{a}o e F\'{\i}sica Experimental de Part\'{\i}culas , 3004-516~Coimbra, Portugal}
\address[c]{Smoluchowski Institute of Physics, Jagiellonian University of Cracow, 30-059~Krak\'{o}w, Poland}
\address[d]{GSI Helmholtzzentrum f\"{u}r Schwerionenforschung GmbH, 64291~Darmstadt, Germany}
\address[e]{Technische Universit\"{a}t Darmstadt, 64289~Darmstadt, Germany}
\address[f]{Institut f\"{u}r Strahlenphysik, Helmholtz-Zentrum Dresden-Rossendorf, 01314~Dresden, Germany}
\address[g]{Joint Institute of Nuclear Research, 141980~Dubna, Russia}
\address[h]{Institut f\"{u}r Kernphysik, Goethe-Universit\"{a}t, 60438 ~Frankfurt, Germany}
\address[i]{Excellence Cluster 'Origin and Structure of the Universe' , 85748~Garching, Germany}
\address[j]{Physik Department E12, Technische Universit\"{a}t M\"{u}nchen, 85748~Garching, Germany}
\address[k]{II.Physikalisches Institut, Justus Liebig Universit\"{a}t Giessen, 35392~Giessen, Germany}
\address[l]{Istituto Nazionale di Fisica Nucleare, Sezione di Milano, 20133~Milano, Italy}
\address[m]{Institute for Nuclear Research, Russian Academy of Science, 117312~Moscow, Russia}
\address[n]{Department of Physics, University of Cyprus, 1678~Nicosia, Cyprus}
\address[o]{Institut de Physique Nucl\'{e}aire (UMR 8608), CNRS/IN2P3 - Universit\'{e} Paris Sud, F-91406~Orsay Cedex, France}
\address[p]{Nuclear Physics Institute, Academy of Sciences of Czech Republic, 25068~Rez, Czech Republic}
\address[r]{LabCAF. Dpto. F\'{\i}sica de Part\'{\i}culas, Univ. de Santiago de Compostela, 15706~Santiago de Compostela, Spain}
\address[s]{Instituto de F\'{\i}sica Corpuscular, Universidad de Valencia-CSIC, 46971~Valencia, Spain}

\author[g]{G.~Agakishiev} 
\author[c]{A.~Balanda}
\author[r]{D.~Belver}
\author[g]{A.V.Belyaev}
\author[b]{A.~Blanco}
\author[j]{M.B\"{o}hmer}
\author[o]{J.L.Boyard}
\author[d]{P.~Braun-Munzinger\fnref{fn3}}
\author[r]{P.~Cabanelas}
\author[r]{E.~Castro}
% ...

\cortext[cor1]{Corresponding authors.}
\fntext[fn1]{also at Lawrence Berkeley National Laboratory, ~Berkeley, USA.}
\fntext[fn2]{also at ISEC Coimbra, ~Coimbra, Portugal.}
\fntext[fn3]{also at ExtreMe Matter Institute EMMI, 64291~Darmstadt, Germany.}
\fntext[fn4]{also at Technische Universit\"{a}t Dresden, 01062~Dresden, Germany.}
\fntext[fn5]{also at Dipartimento di Fisica, Universit\`{a} di Milano, 20133~Milano, Italy.}
\fntext[fn6]{also at Frederick University, 1036~Nicosia, Cyprus.}
\fntext[fn7]{also at Dipartimento di Fisica Generale and INFN, Universit\`{a} di Torino, 10125~Torino, Italy.}

\begin{abstract}
The tagged quasi-free $np \to np\pi^+\pi^-$ reaction 
has been studied experimentally with the High Acceptance Di-Electron Spectrometer (HADES) at GSI at a deuteron 
incident beam energy of 1.25 GeV/nucleon ($\sqrt s \sim$ 2.42 GeV/c for the quasi-free collision).
 For the first time, differential distributions of solid statistics for 
 $\pi^{+}\pi^{-}$ production in $np$ collisions have been collected in the region corresponding to the large transverse momenta of the secondary particles.
The invariant mass and angular distributions for the $np\rightarrow np\pi^{+}\pi^{-}$ reaction are compared with different models. This comparison confirms the dominance of the $t$-channel with $\Delta\Delta$ contribution.
It also validates the changes previously introduced in the Valencia model to describe two-pion production data  in other isospin channels,  although some deviations are observed, especially for the $\pi^{+}\pi^{-}$ invariant mass spectrum. The extracted total cross section is also in much better agreement with this model. Our new measurement puts useful constraints for the existence of the conjectured dibaryon resonance at mass M$\sim$ 2.38 GeV and with width  $\Gamma\sim$ 70 MeV.
\vspace{1pc}

\end{abstract}


\end{frontmatter}

% \linenumbers

\section{Introduction}
\label{introduction}
The two-pion production in nucleon-nucleon ($NN$) collisions is a  rich source of information about the baryon excitation spectrum and the  baryon-baryon interactions. 
In addition to the excitation of a resonance decaying into two pions, which can also be studied in 
the $\pi N \to \pi\pi N$ \cite{piN} and $\gamma N \to \pi\pi N$ \cite{gammaN} reactions, 
 the simultaneous excitation of two 
baryons can be investigated in the NN reactions. By giving access to single and double baryon excitation processes, which both  play an important role in the NN dynamics in the few GeV energy range and  contribute significantly  to   meson  and dilepton production, the two-pion production appears as a key process towards a better understanding of hadronic processes. In comparison to the one-pion decay mode, it presents a different  selectivity with respect to the various  resonances. In particular, the excitation of baryonic resonances coupled to the $\rho$ meson can be studied with the two pions  in the isospin 1  channel. This is of utmost interest for a better understanding of the dilepton production in nucleon-nucleon reactions, where these couplings manifest clearly \cite{HADES_pp22,HADES_pp35_incl,HADES_pp35_excl}, and also in nucleon matter due to the expected  modifications of the  $\rho$ meson spectral functions \cite{Leupold2010}. Finally, the comparison of two-pion production in $pp$ and $np$ channels could shed some light on the origin of the surprisingly large isospin dependence of the dilepton emission observed by the HADES experiment \cite{kirill}. In particular, the $\rho$ production mechanism via \Del \Del\ final state interaction, which does not contribute in the $pp$ channel, was recently proposed as an explanation for the different dilepton yield measured in $pp$ and $pn$ channels \cite{Bashkanov_dileptons13}. It is therefore important to check the description of the double \Del\ process in the two-pion production channels.\par 
Additionally, following  the intriguing results obtained by the WASA collaboration  in the double pionic fusion reactions, a renewed interest on the study of the two-pion production in NN collisions was sparked, in order to check the possible contribution of a dibaryon resonance \cite{ABC-bashkanov,ABC-wasa-cosy-1}.  \par
The answer to all these open questions   requires    
systematic two-pion production measurements   both in proton-proton and neutron-proton collisions.  
Concerning proton-proton collisions, a significant amount of data has  been accumulated for various two-pion final channels in bubble chamber experiments 
\cite{pickup,hart_2.85,eisner,brunt,cochran,cverna,shimizu,dakhno1,dakhno2,lehar,kek1,pnpi1,pnpi2}  for proton incident energies from  the threshold up to 2.85~GeV. Precise differential cross-sections have also been  obtained recently  at CELSIUS and COSY up to 1.4 GeV  \cite{brodowski,johanson,patzold,cosy-tof1,cosy-tof2,ABC-anke,wasa-0_lowenergy,wasa-1_isospin,wasa-2_modVal,wasa-3_nnpi+pi+,wasa-cosy-4_pp14}, with an emphasis on the $\pi^0\pi^0$ production.  The data base for the $pn$ reaction from the bubble chamber experiments is even more scarce \cite{brunt,dakhno1,dakhno2,kek1,dubna}. Very recently, however, precise measurements of total and differential cross sections for the $np \to pp\pi ^{-} \pi^{0}$ and $np \to np\pi ^{0} \pi^{0}$ became available from WASA at COSY at neutron  energies from 1.075 to 1.36 GeV  \cite{wasa-cosy2013_pnpppi-pi0,wasa2015-pi0pi0}. In the $np \to np\pi^+\pi^-$ channel, differential cross-sections  are also known from Dubna measurements \cite{dubna1,jerus2015}, covering  the beam incident energy  range from 0.624 to 4.346 GeV.\par
  
Since the chiral perturbation theory calculations for two-pion production in $NN$ collisions are  
available only near threshold \cite{ChPT}, 
several phenomenological models
have been suggested for the analysis of the double pion production in 
$NN$ collisions in the GeV energy range. 
The first theoretical developments related to the two-pion production were based on the one-pion exchange (OPE) model \cite{Ferrari63}. The reggeized $\pi$ exchange  model (OPER) \cite{oper,oper2}, which uses the partial wave analysis results for $\pi N$ elastic scattering \cite{pwa}, constitutes its most recent and most elaborate modification.  
Lagrangian models were also introduced. The Valencia model   by Alvarez-Ruso et al. \cite{valencia} was first developped. It  aimed at a description of NN collisions at energies lower than 1.4 GeV and included $N(1440)$ and $\Delta$(1232) excitations. The Cao et al.  model \cite{china}, developped after the publication of new data at COSY and CELSIUS, has a larger range of applicability due to the inclusion of resonances with mass up to 1.72 GeV.   Both models qualitatively reproduce the very fast increase of the cross section above threshold in the different two-pion production channels and predict
the dominance of two processes above 1 GeV: the excitation of the   N$(1440)$ resonance and subsequent decay into  \Del$\pi$ or $N\sigma$ and the double $\Delta$ excitation. \par
However, both models \cite{valencia,china} have failed to reproduce the $\pi^0\pi^0$ spectra for the  
$pp\rightarrow pp\pi^{0}\pi^{0}$ reaction at  beam energies above 1.0 GeV \cite{wasa-2_modVal,wasa-cosy-4_pp14}, which motivated the development of the so-called  "modified Valencia model"  \cite{wasa-2_modVal}, providing  a much improved description of these data. This new model  has been  used by the WASA collaboration for the interpretation of the double pionic fusion reactions, after some additional changes to take into account the deuteron formation. However, the observed resonant behavior of  the cross section of the $pn \to d\pi^0\pi^0$   \cite{ABC-bashkanov,ABC-wasa-cosy-1}, associated with a structure at low $\pi^0\pi^0$ invariant mass (the so-called ABC effect) could not  be explained by such an approach and were interpreted as being due to a dibaryon resonance in the I=0 NN channel, with a mass of 2.37 GeV/c$^2$ and a width of 70 MeV. This hypothesis was further supported by the isospin decomposition of the $ pn \to d \pi\pi$ reaction \cite{ABC-wasa-cosy-2_isospin_fusion}. The latter provided a consistent description of both I=0 and I=1 channels by taking into account the resonant contribution in addition to the conventional t-channel processes described by the "modified Valencia model". Even more recently, the $pn \to pp \pi^0\pi^-$ \cite{wasa-cosy2013_pnpppi-pi0} reaction was also consistently described with the same model. 
The accuracy of $d^{*}$ resonance hypothesis is also supported by the SAID partial wave analysis based on new polarized $\vec{n}p$ scattering data.
In this analysis a resonance pole in the $^{3}D_{3}-^{3}G_{3}$ coupled partial waves at (2380$\pm$10-i40$\pm$5) MeV have been discovered \cite{PWA-1,PWA-2}.
\par
Considering the impact that  the discovery of a dibaryon resonance could have,  this systematic study must be pursued.  It is indeed important to  
 provide, possibly in independent experiments, constraints for all possible channels of the $pn$ reaction, where the resonance is expected to contribute, in particular \pip\pim\ or \piz\piz\ production with an unbound $pn$ pair, but also for $pp$ channels in order to check unambiguously the consistent description of the conventional processes. \par 
 The experiments with HADES \cite{HADES} are particularly well suited to answer such a request, since both $pp$ and $pn$ experiments were studied in the relevant energy range. The good capacities of  HADES  for hadron identification were already shown in the study of one-pion and one-eta (measured by its three pion decay) production channels  \cite{HADES_pp35_excl,HADES_pi_eta_excl} of the $pp$ reaction at different energies.  Beyond the search for the dibaryon resonance, the study of the two-pion production process by the HADES collaboration is also motivated by the connection with the dilepton production, as mentioned above. \par

As a first step in this program, we report here on  the analysis of the tagged quasi-free $pn \to pn \pi^+\pi^-$ reaction from experiments using a deuteron beam of 1.25 GeV/nucleon and present precise total and differential cross-sections. 
 According to previous estimates \cite{brunt,dakhno1,dakhno2,kek1}, this channel is dominated by the isospin-0 contributions, in which the resonance should reveal. In the present analysis, the results were averaged over the available range of $np$ center-of-mass energies.  Our strategy consists in comparing the various differential spectra to three different models introduced above: the OPER model, the modified Valencia and the Cao model and check whether  a description of the data is possible without contribution from the dibaryon resonance. 
The sensitivity of our data to the different mechanisms (double $\Delta(1232)$, $N(1440)$ excitation, as well as higher lying resonances)  can be studied.  The total cross-section can be used for  consistency checks of the different analysis,  in comparison to the already measured $pn \to d\pi^+\pi^-$ cross-section, as  discussed in \cite{Faldt2011,Albaladejo2013}.\par 

Our paper is organized as follows: The experimental procedure is described in Sec.~\ref{experiment}. The features of the models used for the description of the data are introduced in Sec.~\ref{models}. We present and discuss the experimental results and the comparison with the models in Sec.~\ref{results} and draw conclusions in Sec.~\ref{conclusion}.

\section{Experimental procedure}
\label{experiment}
The experimental data have been obtained using HADES \cite{HADES}  located
at the GSI Helmholtzzentrum f\"ur Schwerionenforschung
in Darmstadt, Germany.
 HADES is a modern multi-purpose detector currently operating in the region of kinetic beam energies of up to 2~A$\cdot$GeV for nucleus-nucleus collisions.
The detailed description of the set-up can be found in \cite{HADES}, here we briefly summarize the main features relevant for the present analysis.

HADES is divided into 6 identical sectors 
defined by the superconducting coils producing the toroidal geometry magnetic field.
The spectrometer has 85\% of azimuthal acceptance and covers polar angles from 18$^\circ$ to 85$^\circ$ measured relatively the beam direction. 
Each sector of the spectrometer contains a Ring Imaging Cherenkov detector (RICH) operating in the magnetic field-free region, 4 planes of the
Multi-wire Drift Chambers (MDC) located before and after the magnetic field region,
two plastic scintillator walls for the polar angles larger (TOF) and smaller (TOFINO) 
than 45$^\circ$, respectively, and an electromagnetic cascade detector (Pre-Shower) behind TOFINO for particle identification. 
A two-stage trigger system is employed to select events within a predefined charged particle multiplicity interval, as well as the electron candidates.
The investigation of the quasi-free $np$ reactions with the deuteron beam is performed by using a Forward Wall (FW) scintillator hodoscope by registering the spectator protons.
The FW is an array which consists of nearly 300 scintillating cells each 2.54 cm thick. During the $dp$ experiment the FW was located 7 meters downstream the target covering polar angles from 0.33$^\circ$ up to 7.17$^\circ$.
A Monte Carlo simulation for deuteron-proton breakup has shown that approximately 90\% of all spectator protons are within the FW acceptance \cite{kirill}.

In the present experiment, the deuteron beam with intensity up to $10^7$ particles/s and 1.25 GeV/u of kinetic energy was directed onto a 5 cm long liquid-hydrogen target of 1\% interaction probability. 
The analysis of the deuteron induced quasi-free $np$ reactions 
was based on the first-level triggered events, with the required 
charged particle multiplicity of at least three  in TOF and TOFINO  and a signal  
in the FW hodoscope.  
The information from RICH, TOF/TOFINO and Pre-Shower detectors was used to 
select and discriminate the hadrons against the leptons.
The momentum reconstruction was carried out by measuring the deflection angle
of the particle trajectories derived from the four hit positions in the   
MDC planes (two before and two after the magnetic field zone) using the Runge-Kutta algorithm \cite{HADES}.
The achieved momentum resolution was 2-3\% for protons and pions depending on their momentum and 
scattering angle.

%Quasi-free $np$- interactions have been selected by the detection of the proton spectators with the momenta between 1.6~GeV/$c$ and 2.6~GeV/$c$ reconstructed from the time-of-flight measurement in FW \cite{kirill}.
{Quasi-free $np$ interactions have been selected by the detection of the proton spectators
with scattering angles $\le$2$^\circ$ and  momenta between 1.7~GeV/$c$ and 2.3~GeV/$c$ reconstructed from the  time-of-flight measurement in FW. The latter window is centered on half of the beam momentum and has a half width equals to three times the width of the momentum resolution.} 
The $np \to np\pi^{+}\pi^{-}$ channel selection was based on identification of three detected hadrons
where one of them has the negative momentum polarity. % stop S.V. first 7 pages  
The particle identification was based on event hypotheses where any of the three selected hadrons has been used as the reference particle. 
The reference particle time-of-flight was calculated using  the reconstructed 
momentum and trajectory length. The velocities of the other two particles were then deduced, 
using only the time-of-flight difference with the reference particle.
The applicability of this time-of-flight reconstruction algorithm was checked in a dedicated experiment with a low beam intensity by means of the START detector as discussed in \cite{HADES}.
The correlations between the velocity and reconstructed momentum for all three particles were taken into consideration to reject the wrong hypotheses. 
The correlation  between the energy losses in MDCs and momentum was
additionally applied for the final selection of the protons and $\pi^+$.
{Finally, the total number of selected events corresponding to the $np \to np\pi^{+}\pi^{-}$ channel was  $\sim 8\cdot10^{5}$.  }

\begin{figure}[h]
\includegraphics[width=90mm,height=65mm]{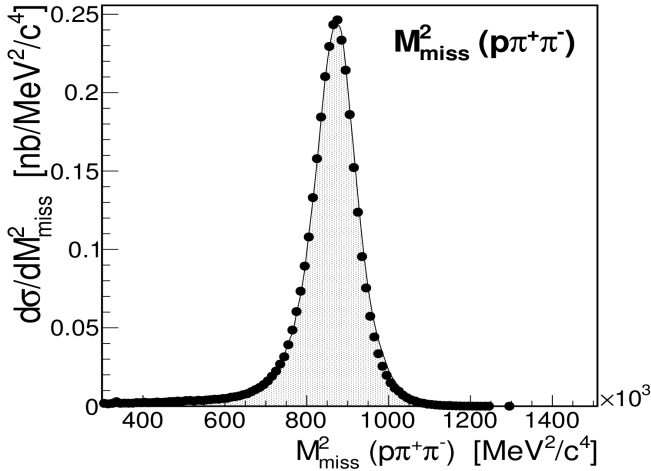}% Here is how to import EPS art
\caption{\label{M2_miss}  
Squared missing mass distributions of the $p\pi^{+}\pi^{-}$ system for the $np \to np\pi^+\pi^-$ reaction at 1.25 GeV. The experimental data are shown by the solid symbols. The shaded area
displays  the phase-space distributions at 1.25 GeV, corrected for the energy dependence of total cross section taken from \cite{lehar}.}
\end{figure}

The normalization of the experimental yield has been performed using the simultaneously measured quasi-elastic $pp$ scattering yield \cite{EDDA}.
The selection of $pp$ elastic events was based on relations between the polar angles $\theta_{1}$ and $\theta_{2}$ and azimuthal angles $\phi_{1}$ and $\phi_{2}$ of both protons due to the momentum conservation in
elastic scattering : 
\begin{eqnarray}
|\phi_{1}-\phi_{2}| = 180^{\circ},
\end{eqnarray}
\begin{eqnarray}
\tan\theta_{1}\cdot \tan \theta_{2} = \frac{1}{\gamma_{CM}^{2}},
\end{eqnarray}
where $\gamma_{CM}$ is the Lorentz factor of the center-of-mass  (CM) system. The elastic events were approximately selected by an elliptic cut in
the $|\phi_{1}-\phi_{2}|$ vs. $\tan\theta_{1}\cdot\tan\theta_{2})$ plane with semi axes corresponding to  $3 \sigma$ cut for each variable.
The quasi-free $pp$ elastic data measured by HADES in the angular range $46^{\circ} - 134^{\circ}$ in the CM system have been corrected for the efficiency and acceptance 
by using a Pluto \cite{pluto1,pluto2} simulation which uses the data from \cite{EDDA} as an input and takes into account the proton momentum distribution in the deuteron and the dependence of the cross section on the $pp$ center of mass energy dependence of the cross section.
The resulting factor taken to normalize the differential cross section has a precision of about 4\% at 1.25 GeV. 
The details of the normalization procedure at 1.25~ GeV are discussed in \cite{HADES_pi12}.

{ The purity of the $np \to np\pi^{+}\pi^{-}$ channel selection is demonstrated by the squared missing mass distribution of $p\pi^{+}\pi^{-}$ shown as solid circles in Fig.~\ref{M2_miss}, which peaks close to the neutron mass. The shaded area displays the result of a simulation of the quasi-free $np \to np\pi^{+}\pi^{-}$ reaction, where the neutron momentum   distribution in the deuteron is taken into account using the Paris potential \cite{paris} and the $\pi^{+}\pi^{-}$  production in the $n+p$ reaction is treated using phase-distributions and considering a rise of the  cross section with the $np$ center-of-mass energy according to \cite{lehar}.

The direct comparison of  theoretical and efficiency corrected experimental distributions has been performed inside the HADES acceptance using  dedicated filters, as will be shown in the following.
For this purpose, the acceptances and efficiencies for different particles (i.e. pions and protons) were separately tabulated in matrices as  functions of the momentum, azimuthal and polar angles.
The acceptance matrices describe only the HADES fiducial volume and can be applied as a filter to the events generated by the models. 
The matrix coefficients have been determined by means of full GEANT simulations, with pions and protons processed through the detector and reconstructed by tracking, particle identification and selection
algorithms with the same package as it was done for real events. 
The acceptance of HADES for the $np \rightarrow np\pi^{+}\pi^{-}$ reaction is about $6\%$ in comparison with the full phase-space.
The resulting detection and reconstruction efficiency is typically about 90\% for protons and pions.
These efficiencies have a very smooth behavior as a function of the different observables shown in the following. Therefore, we only consider a global systematic uncertainty which will affect the normalization of our data, but not the shape of the spectra. This effect is partially taken into account by the normalization to the elastic data, and we estimate the residual uncertainty to 2$\%$.  

\section{Model features}
\label{models}
For the interpretation and discussion of our results in sec.~\ref{results} below, we  used the modified Valencia  \cite{wasa-3_nnpi+pi+}, the  Cao \cite{china} and OPER models \cite{oper2}. 
The Cao and modified Valencia models are effective lagrangian models taking into account different resonant and non-resonant graphs. The original Valencia model \cite{valencia} includes only $N(1440)$ and $\Delta(1232)$ resonances while the Cao model considers all known baryonic resonances with mass up to 1.72~GeV,  neglecting the interferences between the different contributions, included in the Valencia model. The practical differences in our energy range are lower contributions for both  \Del\ and \nstar\ by about 30$\%$ in the Valencia model, which are partially compensated by constructive interferences.
 In comparison with the original Valencia model, four main changes have been applied in the "modified" Valencia model. 
First, the ratio R= $\Gamma(N(1440) \to \Delta\pi)/\Gamma(N(1440) \to N\sigma$) of the branching ratios of the  Roper resonance towards 2$\pi$N via $\Delta\pi$ or $N\sigma$ has been reduced from 4 in the initial model to 1. The initial value corresponded to PDG \cite{pdg} estimates prior to 2012, and a new value was deduced from an analysis of the $\pi^0\pi^0$ opening angle and invariant mass distributions obtained in the $pp \to pp\pi^0\pi^0$  reaction below 900 MeV \cite{wasa-0_lowenergy} assuming the dominance of the Roper excitation.  This new value is also in agreement with a recent Partial Wave Analysis \cite{pwa-sar-1,Anisovich:2011fc}. In the meantime,   the PDG limits for the ratio have also been changed and are now between 1 and 3. This change seems therefore fully consistent. The Cao model uses a value 2 for this ratio, hence favoring $N(1440) \to \Delta\pi$  by a factor 2 with respect to  $N(1440) \to N\sigma$ decay, while they have equal weights in the modified Valencia model. \par
The second change is a readjustment of the $N(1440)$ strength in the modified Valencia model according to the isospin decomposition of the two-pion production channels in the $pp$ reaction  at different energies between 0.775 GeV and 1.36 GeV \cite{wasa-1_isospin}. The maximum reduction is obtained exactly for our incident energy of 1.25 GeV and amounts to a factor 2. 
%This brings the $N(1440)$ contribution to be much smaller than the double $\Delta$ excitation in the modified Valencia model and makes a significant difference with the Cao model, where  the $N(1440)$ contribution corresponds to about 40~$\%$ of the total $pn \to pn\pi^+\pi^-$ cross section at an incident energy of 1.25 GeV.
This brings the $N(1440)$ contribution to be much smaller than the double $\Delta$ excitation in the modified Valencia model and makes a significant difference with the Cao model, where  the $N(1440)$ contribution corresponds to about 40~$\%$ of the total $pn \to pn\pi^+\pi^-$ cross section at an incident energy of 1.25 GeV.\par 
The third modification is driven by the shape of the $\pi^0\pi^0$ invariant mass distribution measured in  $pp \to pp\pi^0\pi^0$ at an incident energy larger than 1 GeV, where the double $\Delta$ mechanism dominates \cite{wasa-2_modVal}. It consists in a reduction  by a factor 12, and a change of the sign of the $\rho$ exchange contribution in the $\Delta$ excitation mechanism.  The latter indeed induced a two-hump structure in the $\pi^0\pi^0$ invariant mass distribution, which was not seen, neither in the data \cite{wasa-2_modVal} nor in the Cao model which has a very small $\rho$ exchange contribution. \par
The fourth modification consists in  the introduction of the \Del (1600) $\to$ \Del (1232)$\pi$ contribution to improve the description of the \ppnnpippip\ reaction. In this channel, the N* decay does not contribute, and the cross section is larger by more than a factor 2 than expected for the double \Del (1232) contribution only. The WASA collaboration explained this feature by a large contribution of  \Del (1600) excitation, with a destructive interference with the double $\Delta$(1232) contribution. After introducing this new contribution, the \ppnnpippip\ cross section could be reproduced and the description of the differential spectra, especially the \pip\pim\ opening angle and invariant mass distributions measured for this channel at 1.1 GeV \cite{wasa-3_nnpi+pi+},          improved significantly. The \Del (1600) $\to$ \Del (1232)$\pi$ process was implemented in a similar way as  \Del (1232) $\to$ \Del (1232)$\pi$, which was already taken into account in the Valencia model. However, such a large contribution of the $\Delta(1600)$ resonance at such low energies is surprising. In the Cao model, it is taken into account, but starts to contribute significantly to the \ppnnpippip\ reaction for energies larger than 1.6 GeV. It has to be noted that, while the original Valencia model underestimates the \ppnnpippip\ cross section by more than a factor 2, the Cao model provides good predictions for both total cross sections  and differential spectra \cite{china}. This is related to the larger double $\Delta$ contribution in this model. However, such a large double \Del\ contribution leads to a significant overestimate of the $pp \to pp$\piz\piz\ cross section. In both models, the  nucleon pole terms are found to be very small, although they have a much larger relative contribution in the \ppnnpippip\  than in other channels.  \par 

  As mentioned in the introduction, the OPER model is based on  a Reggeized $\pi$ -exchange model and uses    
on-shell amplitudes of the elastic $\pi N$ scattering and of the inelastic $\pi N \to \pi\pi N$ reaction, with form factors and propagators taking into account the off-shellness of the exchanged pion. The elastic amplitudes are taken from a Partial Wave Analysis and the inelastic ones are deduced from a parametrization obtained in the framework of the Generalized Isobar Model \cite{dubna1,Manley}. This model provided a good description of differential spectra measured at Dubna in the $np \to np$\pip\pim\ reaction at 5.2 GeV/c and in  $\bar{p}p \to \bar{p}p$\pip\pim\ at 7.23 GeV/c \cite{dubna1}. The One-Baryon Exchange (OBE) diagrams were introduced, as described in \cite{dubna1}, to improve the description of the $np \to np$\pip\pim\ spectra below 3 GeV/c. Such diagrams correspond  in Lagrangian models  to the  "pre-emission contribution". Very recently,  the contribution of the so called "hanged" diagrams, due to the two-pion production from the exchange pion line, was also considered. For this, amplitudes for   $\pi\pi$ scattering were used. This addition  was shown to improve the description of the $\pi\pi$ invariant mass spectra in the low mass region \cite{oper2}. Such graphs are taken into account in the Valencia model, but are neglected in the Cao model.   Table \ref{tab:contributions} displays the main resonant contributions in the three models at an incident neutron energy of 1.25 GeV, disregarding the contribution via interference terms.  This makes a big difference  with the OPER model, where the OBE diagrams, where at least one pion is not emitted by a resonance, amount to 41$\%$ of the total yield.  Although these numbers do not take into account interference effects, they already point to major differences between the models which will be investigated further when comparing to our data. \par

\begin{table}[bh!]
\begin{tabular}{|l|l|l|l|}
\hline
& Cao &  mod. Valencia &OPER \\
\hline
\Del (1232)\Del (1232) & 47.0 $\%$& 60 $\%$ & 38.0 $\%$ \\ 
\hline
$N(1440) \to \Delta (1232)\pi$  & 23.0 $\%$  & 2.1 $\%$ & 4.5 $\%$ \\
\hline
$N(1440) \to N\sigma$   & 20.0 $\%$ & 8.2 $\%$ & 0.2 $\%$  \\
\hline	
 \Del (1600)$\to\Delta (1232)\pi$  &  3.0 $\%$& 21.0 $\%$ & 4.5 $\%$ \\
\hline
\end{tabular}
\caption{Main contributions  in the Cao \cite{china}, "modified Valencia" \cite{
wasa-2_modVal} and OPER \cite{oper2} models for the \pnpnpippim\ for a neutron incident energy of 1.25 GeV.}
\label{tab:contributions}
\end{table}
To  take into account the momentum distribution of the neutron inside the deuteron, 
 the Paris \cite{paris} deuteron wave functions has been used
 for the phase-space calculation, the Hulthen \cite{hulthen} wave function for  modified Valencia \cite{wasa-3_nnpi+pi+}  and CD-Bonn \cite{cdbonn} for OPER \cite{oper2} and  Cao \cite{china} models. No significant difference is expected from these  different inputs. As mentioned in the previous section, events have been generated according to  distributions from these models  and filtered by the HADES acceptance.

\section{Experimental results and comparison with models}
\label{results}
We first concentrate on  the comparison of the shapes of the theoretical and experimental distributions and will discuss the cross sections at the end of this section.
The invariant mass and angular distributions for the $np\rightarrow np\pi^{+}\pi^{-}$ reaction at 1.25 GeV measured inside the HADES acceptance and corrected for the reconstruction efficiency are presented 
by solid circles in Figs.~\ref{np125_mass}-\ref{np125_contributions_mass} and \ref{np125_angular}-\ref{np125_contributions_angle} respectively.  
Only statistical errors are shown. As explained above, the global uncertainty taking into account both normalization and efficiency correction is of the order of 5$\%$, with no significant dependence on the different observables. \par

 The predictions from the modified Valencia \cite{wasa-3_nnpi+pi+}, Cao  \cite{china} and OPER models\cite{oper2}  have been normalized therefore  to the total experimental yield and are presented  as  long-dashed,  dashed  and solid lines, respectively, in Figs.~\ref{np125_mass} and ~\ref{np125_angular}. The hatched areas correspond to  phase-space distributions obtained in the same way. In addition, the different contributions in the OPER and modified Valencia models are shown in Figs.~\ref{np125_contributions_mass} and \ref{np125_contributions_angle} in comparison to selected experimental distributions exhibited in Figs.~\ref{np125_mass} and \ref{np125_angular}.
 \subsection{Invariant mass distributions}
\label{InvMass}
\begin{figure}[h]
\includegraphics[width=90mm,height=80mm]{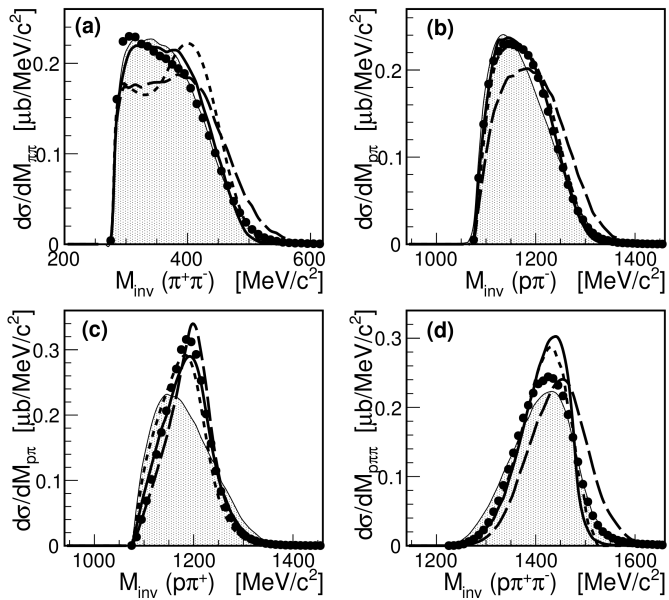}% Here is how to import EPS art
\caption{\label{np125_mass}  
Distributions of the $\pi^{+}\pi^{-}$ (a), $p\pi^{-}$ (b), $p\pi^{+}$ (c) and $p\pi^{+}\pi^{-}$ (d) invariant masses
for the $np \to np\pi^+\pi^-$ reaction at 1.25~GeV. The experimental data are shown by solid symbols. The theoretical predictions within  HADES acceptance from OPER \cite{oper2}, Cao \cite{china} and modified Valencia models \cite{wasa-3_nnpi+pi+} are given by the solid,  dashed and  long-dashed curves, respectively. The shaded areas show the phase-space distributions.}
\end{figure}

\begin{figure}[ht]
\begin{center}
\includegraphics[width=90mm,height=120mm]{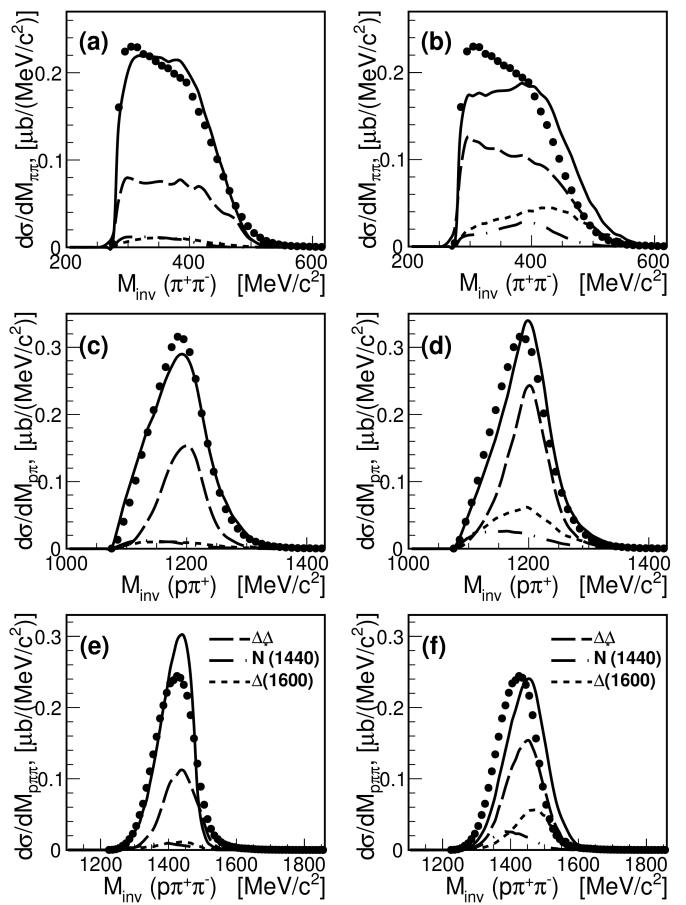}% Here is how to import EPS art
\caption{\label{np125_contributions_mass}
Experimental data (full dots) for  \pnpnpippim\ at 1.25 GeV are compared to the total yield (solid curves) for the OPER model \cite{oper2} (left column) and modified Valencia model \cite{wasa-2_modVal,wasa-3_nnpi+pi+} (right column). In each case, the \Del\Del\ (long-dashed), $N(1400)$ (long dash-dotted) and \Del (1600) (short-dashed) contributions are shown.
a) and b): $\pi^{+}\pi^{-}$ invariant mass , c) and d): $p\pi^{+}$ invariant mass, e) and f):  $p\pi^{+}\pi^{-}$ invariant mass distributions.
}
\end{center}
\end{figure}

The experimental $p\pi^{-}$, $p\pi^{+}$ and $p\pi^{+}\pi^{-}$ invariant mass distributions (panels $b)$, $c)$ and $d)$ in Fig.~\ref{np125_mass}) are all shifted to  higher masses in comparison with the 
phase-space calculations. In particular, the invariant mass distribution of the  $p\pi^{+}$ subsystem (panel c) in Fig.~\ref{np125_mass}) shows a pronounced resonant behavior with a position of the maximum in the experimental distribution roughly corresponding to the $\Delta^{++}$ mass. This  distribution is  well described by the different models and deviates significantly from the phase space distribution. 
Such a resonant behavior of the p$\pi^{+}$ invariant mass distribution is expected for the double \Del\ excitation process, since the isospin factors favor   the excitation of the $\Delta^{++}\Delta^{-}$ contribution with respect to the $\Delta^{+}\Delta^{0}$  by a weight of 8/5. A resonance in the $p$\pip\ system is also expected for the $N(1440)^+$ decay into \Del$^{++}\pi^{-}$, but with a "smearing" due to the  excitation of the $N(1440)^0$ and subsequent decay into \Del$^{-}\pi^{+}$ which has the same probability. In addition, the dominance with respect to the other isospin channels is also lower than in the case of the double \Del\ excitation. The situation  is similar for the \Del (1600) case, with an even lower relative weight for the decay involving the \Del (1232)$^{++}$ excitation. These statements are confirmed by the behavior of the $N(1440)$ and \Del(1600) contributions which are displayed  in Fig.~\ref{np125_contributions_mass}d below in the case of the modified Valencia model.  The overall contribution of the double \Del\ excitation is larger in the modified Valencia model than in the Cao model. This is probably the reason why the latter presents a slightly broader $p\pi^{+}$ invariant mass distribution. Surprisingly, the $p$\pip\ invariant mass of the double \Del (1232) contribution in the OPER model is slightly broader than in the Valencia model, as can be seen in Fig.~\ref{np125_contributions_mass}c. The global distribution which is further broadened by the OBE contributions, is similar to the one in the Cao model with an additional shift to higher energies (see Fig.~\ref{np125_mass}b).     \par
 The deviations from phase-space distributions are less spectacular for the $p$\pim\ and $p$\pip\pim\ invariant mass distributions (panels b) and d) in Fig.~\ref{np125_mass}). No clear resonance behavior is indeed expected for the $p$\pim\ system, since   the \Del (1232)$^0$ is  disfavored by isospin with respect to \Del$^{++}$  in all channels (double \Del ,  $N\to$\Del (1232)$\pi$, and the \Del(1600)$\to$\Del(1232)$\pi$). This also holds for the $p$\pip\pim\ system, due to the strong double $\Delta$ contribution. In addition, the $N(1440)^0$ excitation, which is as probable as the $N(1440)^+$, is favored by acceptance. This can be checked in Figs.~\ref{np125_contributions_mass}c and~\ref{np125_contributions_mass}d below, where the $N(1440)$ contributions are indeed shifted to low $p$\pip\pim\ invariant mass.  For the \Del(1600) excitation, the positive charge state, decaying into $p$\pip\pim , is favored by acceptance. This is probably the reason why the $p$\pip \pim\ invariant mass distribution for this contribution is shifted in the models to the high-mass part of the available phase space. The OPER and Cao models give similar predictions for the $p$\pim\ and $p$\pip\pim\ observables and achieve a reasonable description of the  $p$\pim\ but predict a too narrow $p$\pip\pim\ invariant mass distribution. In the case of the Valencia model,  both the $p$\pim\ and  $p$\pip\pim\ are overestimated on the high energy side. This  points to a too large \Del (1600) contribution, as will be discussed below in some details.\par
The three models present the largest deviations for the  $\pi^{+}\pi^{-}$ invariant mass distribution. The sensitivity of this distribution to the two-pion production mechanism in $NN\to NN \pi\pi$ reactions has already been demonstrated in previous work \cite{china,wasa-0_lowenergy,wasa-2_modVal,wasa-cosy-4_pp14}. In fact, the puzzling enhancement observed in the low-energy part of the invariant mass spectra of two pions produced in $pn$ or $pd$ fusion reaction  (the so-called ABC effect) was also a manifestation of this sensitivity and triggered the onset of double \Del\ excitation models \cite{Risser}.   The OPER model is not too far from the data, but has  a slightly too flat shape, while the Cao and Valencia models fail strongly to describe the shape of this distribution. 
The double-hump structure which can be seen in the predictions of the Cao model  is completely absent in our data.
A very similar trend  was already observed for the comparison of the Cao model predictions to the spectra measured in the $pp \to pp$\pip\pim\ and $pp \to pp$\piz\piz\ reactions above 1.1  GeV  \cite{china}.  Since this double-hump structure is due to  the \nstar $\to$ \Del $\pi$ process, these results indicate that this contribution is too large in the Cao model. The OPER model gives the best description of the low mass part of this distribution, which seems to be mainly due to the addition of the OBE and "hanged" diagrams. However, the double \Del\ contribution is also much broader than in the case of the modified Valencia model (see Figs.~\ref{np125_contributions_mass}a and b below). In the modified Valencia model, the excess in the region above 350 MeV/c$^2$ is due to both the \Del (1600) and the Roper contributions and their interference with the double \Del(1232) contribution. In particular, the interference between the Roper and double \Del (1232) contribution is small in overall, but has a significant constructive effect of the level of 10-15$\%$ in this region. Our data would be better described by a change of sign of this interference together with a reduction of the \Del(1600) contribution. 
\subsection{Angular distributions}
\label{AngDistr}

\begin{figure}[h]
\includegraphics[width=90mm]{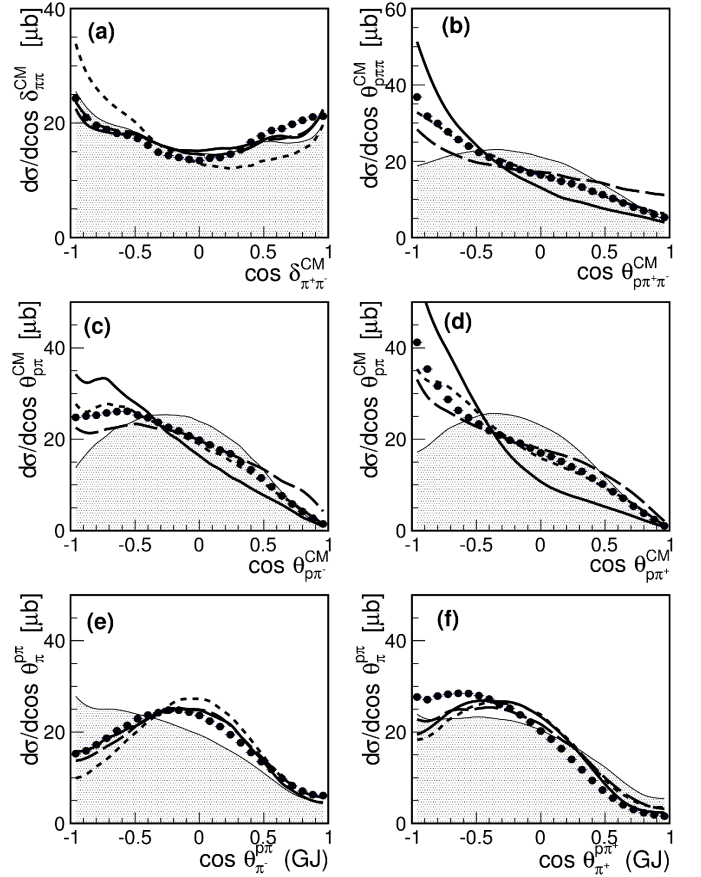}% Here is how to import EPS art
\caption{\label{np125_angular} Same as Fig.~\ref{np125_mass}, but for angular distributions: 
a) - opening angle of $\pi^{+}\pi^{-}$ in the $np$ rest frame, 
b) - polar angle of $p\pi^{+}\pi^{-}$ in the $np$ rest frame, c) - polar angle of $p\pi^{-}$ in the $np$ rest frame, d) - polar angle of $p\pi^{+}$ in the $np$ rest frame,
e) - polar angle of $\pi^{-}$ in the $p\pi^{-}$ Gottfried-Jackson frame, f) - polar angle of $\pi^{+}$ in the $p\pi^{+}$ Gottfried-Jackson frame.}
\end{figure}

\begin{figure}[h]
% \begin{figure}[ht]
\begin{center}
\includegraphics[width=90mm]{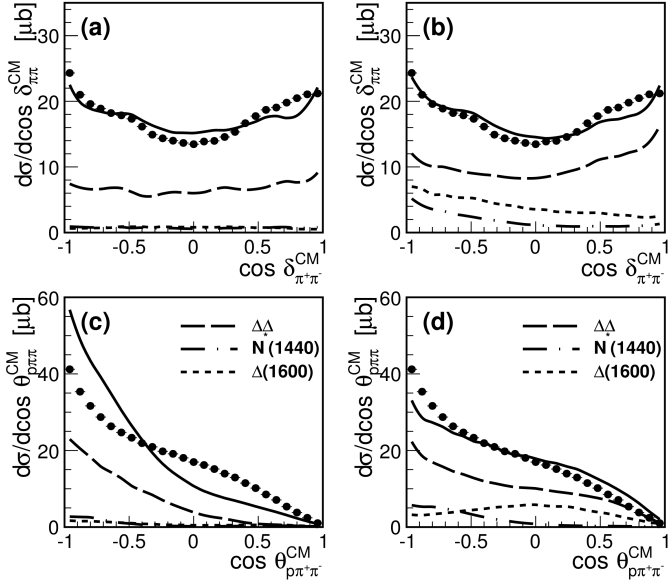}% Here is how to import EPS art
\caption{\label{np125_contributions_angle}
Same as Fig.~\ref{np125_contributions_mass} for 
 opening angle of $\pi^{+}\pi^{-}$ in the $np$ rest frame ( a) and b)) and  
 polar angle of $p\pi^{+}$ system in the $np$ rest frame ( c) and d)).}
\end{center}
\end{figure}
Figure~\ref{np125_angular}   exhibits the angular distributions for the $np\rightarrow np\pi^{+}\pi^{-}$ reaction at 1.25 GeV. 
%{\slshape 
Panels $a)$, $b)$, $c)$, $d)$ and $e)$, $f)$ correspond to the distributions of the opening angle of $\pi^{+}\pi^{-}$, 
polar angles of $p\pi^{-}$, $p\pi^{+}$, $p\pi^{+}\pi^{-}$  subsystems in the $np$ CM system  and polar angles 
of $\pi^ {-}$ and $\pi^{+}$ in the $p\pi^{-}$ and $p\pi^{+}$ Gottfried-Jackson  frames, respectively.
The CM frame was defined assuming the neutron at rest in the deuteron and the angles in the  Gottfried-Jackson frame  are defined with respect to the beam direction. 
%}
%Panels $a)$ , $b)$, $c)$, $d)$ and $e)$, $f)$ correspond to the distributions of the $\pi^{+}\pi^{-}$ opening angle 
%$\delta^{CM}_{\pi^+\pi^-}$, polar angles of $p\pi^{-}$, $p\pi^{+}$, $p\pi^{+}\pi^{-}$ - 
%subsystems in the cms and polar angles 
%of $\pi^{-}$ and $\pi^{+}$ in the $p\pi^{-}$ and $p\pi^{+}$ rest frames ($i.e.$ Gottfried-Jackson angles), respectively.  

 The excess of the measured yields at small $\pi^{+}\pi^{-}$ opening angles 
($\cos\delta^{CM}_{\pi^+\pi^-}\sim$ 1) with respect to the predictions of all models observed in Fig.~\ref{np125_angular} reflects the enhancement at small  $\pi^{+}\pi^{-}$ invariant mass which was observed in Fig.~\ref{np125_mass}a. Both variables are indeed strongly  correlated. The large asymmetry of the distribution in the case of the Cao model and its steep peaking for back to back \pip \pim\ emission  ($\cos\delta^{CM}_{\pi^+\pi^-}=$ -1) are related to the high-mass structure in Fig.~\ref{np125_mass}a, which, as  already mentioned,  is due to the $N(1440) \to$\Del(1232)\ $\pi$ decay. As explained in \cite{wasa-0_lowenergy}, this effect is produced by the double p-wave decay which gives an amplitude with a $\cos^2\delta^{CM}_{\pi^+\pi^-}$ behavior. 
 For the other models, the backward peaking is consistent with the data. It can be checked in Figs.~ \ref{np125_contributions_angle}a and b that the distributions are rather smooth for all contributions. In the case of the modified Valencia model, the roughly symmetric distribution results from the opposite trends of the  double \Del(1232) and \Del (1600) which have yields respectively forward and backward peaked in the HADES acceptance.  For the OPER model, the anisotropy is mainly due to the OBE terms, all other contributions being very flat. Since the $N(1440)$ contribution in the OPER model is dominated by the \Del$\pi$ decay, we were expecting a larger anisotropy, following the arguments given above. In the case of the modified Valencia model, the $N(1440) \to \Delta (1232) \pi$ decay is suppressed with respect to the $\sigma N$ decay (see Table~\ref{tab:contributions}). It is therefore not surprising that the $N(1440)$ contribution does not present a strong anisotropy. The difference of the shapes of the $\cos\delta^{CM}_{\pi^+\pi^-}$ distributions obtained for the double \Del\ contributions in the modified Valencia  and OPER models is consistent with the already mentioned different behavior of the \pip\pim\ invariant masses (Figs.~\ref{np125_contributions_angle}a and b). 

The  $\theta^{CM}_{p\pi^{+}\pi^{-}}$, $\theta^{CM}_{p\pi^-}$ and $\theta^{CM}_{p\pi^+}$ angular distributions shown in panels $b)$  and  $c) $ and $d)$ panels in Fig.~\ref{np125_angular}, respectively, present a significant forward/backward asymmetry. 
%The  modified Valencia model gives the best overall decsription of these distributions. 
This asymmetry reflects the  fact that protons are emitted preferentially backward in the CM, which is mainly due to the strong   \Del $^{++}$\Del$^{-}$ contribution.  This process is more likely for small four-momentum transfers between  the target proton and the \Del$^{++}$ and the beam neutron and the \Del$^{-}$ respectively, hence the  peaking of the protons at backward angles.  Such a strong asymmetry is not present for the other contributions. However, the effect is also enhanced by  the acceptance, as demonstrated by the distribution of the pure phase space. Due to the lacking  detection capabilities for laboratory polar angles lower than 18$^0$, the detection of the proton is indeed more likely at backward center-of-mass angles.   
The Cao model gives the best overall description of the three distributions. %, except the region close to 180$^0$. 
The modified Valencia model is also doing rather well, however forward/backward asymmetry is smaller than in experimental data.
A much worse description of the slopes of $\theta^{CM}_{p\pi^{+}\pi^{-}}$, $\theta^{CM}_{p\pi^-}$ and $\theta^{CM}_{p\pi^+}$ is obtained in OPER model.
%The modified Valencia model gives the best overall agreement of the three distributions. The Cao model is also doing rather well, except for the overestimate of the yields of events with $\theta^{CM}_{p\pi^-}$ close to 180$^0$. 
%A similar effect is observed for the OPER model, but a much worse description of the slopes of the $\theta^{CM}_{p\pi^{+}\pi^{-}}$ and $\theta^{CM}_{p\pi^+}$ is obtained in this model. 
As shown in Figs.~\ref{np125_contributions_angle}c and ~\ref{np125_contributions_angle}d,  the  double \Del\ contribution is much more strongly backward peaked than in the modified Valencia model. However, the OBE and "hanged" contributions,  which are not displayed in the picture, are also responsible for the too steep  total angular distribution. 
One of the possible reasons of such a deviation can be the omission of   
the interference between the "hanged" and other diagrams \cite{oper2}.  The striking difference with the double \Del\ contribution in the modified Valencia model is probably due to the much lower  cut-off parameters in the vertex form factors which induce a much steeper four-momentum transfer dependence.  Worth to note is the strong difference between the $\theta^{CM}_{p\pi^+}$  and $\theta^{CM}_{p\pi^-}$ distributions, in contrast with the  phase space distribution, which allows to appreciate the small effect of the different \pip\ and \pim\ acceptances. 
 Again, the modified Valencia model gives the best simultaneous estimate of these distributions,  which we take as a hint that this model correctly describes the different isospin configurations of the two-pion production mechanisms. It can also be noted that, similarly to  the  \pip\pim\ invariant mass distributions, the description of the $\theta^{CM}_{p\pi^+}$ distribution could be improved by a reduction of the \Del (1600) contribution.
 The distributions of the pion angles $\theta^{p\pi^{-}}_{\pi^{-}} (GJ)$ and $\theta^{p\pi^{+}}_{\pi^{+}} (GJ)$ are shown in panels $e)$ and $f)$ in Fig.~\ref{np125_angular}, respectively.
The quantity $\theta^{ij}_{j}$ denotes the angle between the thee-momentum of  particle $j$ and the direction of the beam  in the center-of-mass of particles $i$ and $j$, taken relatively to the direction of the beam particle in this reference frame. These distributions also present striking differences for \pip\ and \pim . It can be noted that the $\theta^{p\pi^{-}}_{\pi^{-}} (GJ)$ distribution is very well reproduced, especially by the OPER and modified Valencia model, while none of the models predict the observed enhancement for backward  $\theta^{p\pi^{+}}_{\pi^{+}} (GJ)$. \par
To summarize the analysis of these distributions,  the Valencia model provides a much better description than the Cao model. Our analysis therefore validates the changes introduced by the WASA collaboration in the original Valencia model, except perhaps for the $\Delta$(1600) contribution, which seems to be too large. The OPER model, which is based on a very different approach, gives also a good description of the data. In particular, it fits better the \pip \pim\ invariant masses than the modified Valencia model. However, the predictions are worse for the CM angular distributions of the $p$\pip , $p$\pim and $p$\pip \pim\ subsystems.\par

\begin{figure}[h]
	\centering
		\includegraphics[width=90mm]{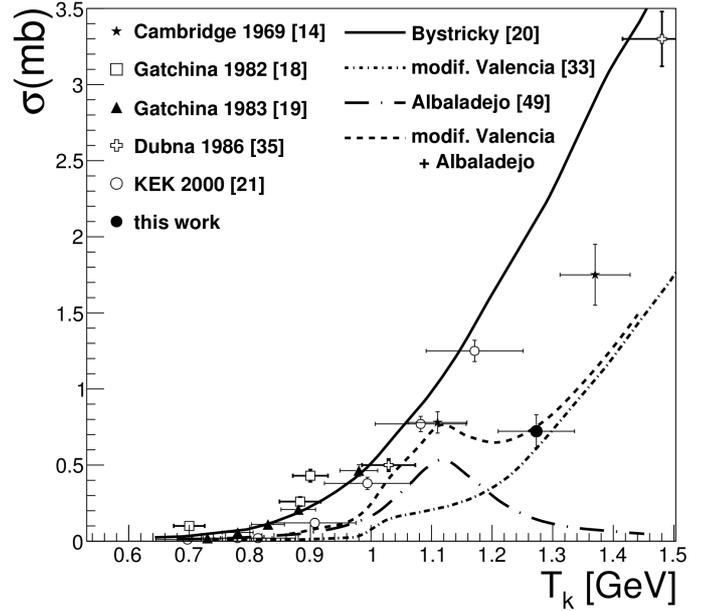} 
	\caption{HADES measurement  for the quasi-free \pnpnpippim\ reaction using a deuterium beam at 1.25 GeV/nucleon (full dot) compared to world data shown by various symbols. The horizontal error bars indicate the spread of the neutron momentum in the different measurements. The full  and short dash-dotted curves  display respectively the "Bistricky parametrization" used for the OPER model \cite{lehar}  normalization and the predictions of the modified Valencia model \cite{wasa-3_nnpi+pi+}. The long dash-dotted curve is the estimate from \cite{Albaladejo2013} for the contribution of the dibaryon resonance. The dashed curve is the sum of the modified Valencia model and dibaryon resonance contributions.}
	\label{fig:crosssection}
\end{figure}
\subsection{Absolute cross sections}
We come now to the discussions of the absolute yields. 
 The measured differential cross section integrated over  the HADES acceptance  is   34.9 $\pm$ 1.5 $\mu$b.  The   model \cite{china} and \cite{oper2} predictions, respectively 72.4 and 86 $\mu $b  are larger by a factor more than 2, while the modified Valencia model \cite{wasa-3_nnpi+pi+}  is in better agreement, with a cross section of 26.4 $\mu$b, i.e., about 30$\%$ lower than the experimental value. Acceptance corrections were calculated using  the modified Valencia model \cite{wasa-3_nnpi+pi+} and OPER model which give a reasonable description
of the differential distributions, as shown in Figs. \ref{np125_mass}-\ref{np125_contributions_angle}. In practice, eight different differential spectra, corresponding to the distributions in $M_{\pi^+\pi^-}$, $M_{p\pi^+}$, $M_{p\pi^-}$, $M_{p \pi^+\pi^-}$, $\cos^{CM}_{\delta_{\pi^+\pi^-}}$,$\cos^{CM}_{p \pi^+\pi^-}$, $\cos\theta^{CM}_{p\pi^+}$ and $\cos\theta^{CM}_{p\pi^-}$ (Figs.~\ref{np125_mass} and \ref{np125_angular} a-d) were corrected for acceptance using both models. The differences between the integrated yields obtained for these different distributions were used to determine the systematic errors. Following this procedure, we obtain for the total cross section  0.65$\pm$0.03 mb  using the modified Valencia model and 0.795$\pm$0.040 mb using the OPER model. Our final estimate, taking into account these two values is $\sigma$= 0.722$\pm$0.108 mb. This cross section is averaged over neutron energies accessible in the quasi-free \pnpnpippim\ reaction at a deuteron beam energy of 1.25 GeV/nucleon. According to the fit of the missing mass (Fig.~\ref{M2_miss}), taking into account the neutron energy and momentum distribution in the deuteron, the  average neutron energy for this measurement is 1.273$\pm$ 0.063 GeV.  Our data point is shown together with the world data in Fig. \ref{fig:crosssection}. In contrast to previous plots which can be found in the literature, we take into account the spread of the neutron momentum in the different measurements. The Cao model  overestimates our point by  a factor 2.4, with a prediction of  1.730 mb, which confirms that this model does not reproduce satisfactorily the \pnpnpippim\ reaction in our energy range. The OPER model does not provide cross sections and was normalized to the "Bistricky parametrization"  \cite{lehar} (shown as a black curve in Fig.~\ref{fig:crosssection}), which resulted from a simple interpolation between measurements over a wide energy range up to  2.2 GeV. This parametrization largely overestimates the NIMROD measurement \cite{brunt}, which was obtained at an incident energy higher by about 120 MeV than  our experiment.  On the other hand, it provided a good prediction for the measurement at KEK \cite{kek1} which was obtained in the meantime.  The latter is of special interest for the present study, since it was obtained for an incident energy only 70 MeV lower than  our experiment. 
Our extrapolated cross section is
approximately a factor 2.6 lower than the "Bistricky parametrization" that has a value of 1.88 mb at 1.25 GeV. This result is in contradiction with a smoothly increasing cross section as a function of the incident energy. This, as we will see in the following, could be expected in the presence of a resonance at lower energies. 
 Our measurement is however hardly compatible with the KEK data ($\sigma=1.25\pm$ 0.05 mb at T$_n$=1.17 GeV). A decrease  of a factor 2 in such a small energy range is indeed difficult to explain.  On the other hand, the modified Valencia  gives a prediction of 590 $\mu$b, i.e. much closer to our value. This model also reasonably reproduces not only the differential distributions for the \pnpnpippim\ channel, as shown in this paper, but also the total cross sections and differential distributions for $pp \to pp$\piz\piz\ \cite{wasa-2_modVal} and $pp \to nn$\pip\pip\ \cite{wasa-3_nnpi+pi+} channels measured by WASA below 1.4 GeV. However, this model does not describe satisfactorily the general excitation function of the \pnpnpippim\ reaction. The underestimation of the data at higher energies might well be due to the lack of higher lying resonances, such as $N(1520)$ and $N(1535)$ in the model. However, another explanation has to be found for the underestimation at lower energies. \par

In the hypothesis of a dibaryon resonance with a mass around  2.38 GeV, as claimed by the WASA collaboration, a structure is expected at a neutron energy around 1.13 GeV.  Using the cross sections for $pn \to d^{\star} \to d\pi^+\pi^-$ extracted by the WASA collaboration, Albaladejo et al. \cite{Albaladejo2013} have estimated the resonant cross section for the \pnpnpippim\ channel, i.e.,  $pn \to d^{\star}\to pn\pi^+\pi^-$ (see dash-dotted curve in Fig.~\ref{fig:crosssection}). They concluded on the  difficulty to reconcile the existing data with the resonance hypothesis, taking into account the large non-resonant contribution which has necessarily to be added.  However, this conclusion is, to our opinion,  strongly biased by the KEK point which is already beyond the resonance and favors a large non-resonant contribution.  In this respect, our measurement, which provides a  much smaller contribution, is in better agreement with the dibaryon resonance hypothesis. The expected resonant contribution at our energy is about 0.2~mb, less than one third of our measured cross section, and is compatible with a non-resonant contribution of the order of 0.5~mb as predicted by the modified Valencia model. According to this model, the non-resonant contribution at the peak of the resonance is only of the order  of 0.2~mb. The existing cross section measurements in this region  are therefore not inconsistent with a resonant cross-section  of about 0.7~mb, as deduced from the WASA result in the $pn \to d$\pip\pim\ channel. To illustrate that, we show in Fig.~\ref{fig:crosssection} the excitation function obtained by adding the dibaryon contribution from \cite{Albaladejo2013} to the modified Valencia model prediction. Except for the KEK point, the result describes within 20$\%$  the cross sections  obtained for incident energies
		between 1 and 1.3 GeV, which is the energy range of the pp$\to$pp\piz\piz\ and pp$\to$nn\pip\pip\ data used to adjust  the modified Valencia model.  In particular, it is in perfect agreement with our measurement. The poorer description outside this energy range could probably be reduced by further adjustments of the Valencia model.  No conclusion can however be drawn from this  very crude calculation, due to the unknown effect of interferences between t-channel  and s-channel processes. In addition, it would be necessary to include the decay of the resonance  into a $pn$ pair with other quantum numbers than the deuteron. According to the analysis in \cite{Bashkanov2015}, this contribution is estimated to be of the order of 0.1 mb. The effect of the resonance contribution on the differential distributions and in particular to the $p\pi^{+}\pi^{-}$ invariant mass is also an open question.   This can only be made using a full model, including in a consistent way the t-channel processes, based on the "modified Valencia" model and the resonant s-channel although a phenomenological approach was presented for the analysis of $np \to  np\pi^{0}\pi^{0}$ in \cite{wasa2015-pi0pi0}.

\subsection{Acceptance corrected distributions}
In order to allow for a more direct comparison of our data to differential cross sections
for quasi-free $np \to np\pi^{+}\pi^{-}$  \cite{jerus2015} and $np \to  np\pi^{0}\pi^{0}$ \cite{wasa2015-pi0pi0} reactions measured in other experiments and possibly facilitate the comparison to other potential models key distributions have been corrected for acceptance.
The correction has been done by using modified Valencia \cite{wasa-3_nnpi+pi+} and OPER \cite{oper2} models.
Fig.~\ref{fig:acc_corr_distr} presents the extrapolated distributions of $\pi^{+}\pi^{-}$ invariant mass and opening angle between pions in CM frame for $np\rightarrow np\pi^{+}\pi^{-}$ reaction at 1.25 GeV. 
%Black points correspond to the experimental data extrapolated in $4\pi$ region.
The errors on extrapolated data take into account  the difference between acceptance correction coefficients for modified Valencia \cite{wasa-3_nnpi+pi+} and OPER \cite{oper2} models.
The predictions of the different models in $4\pi$ are also displayed in Fig.~\ref{fig:acc_corr_distr} using the same normalisation as for the comparison inside HADES acceptance (Fig.~\ref{np125_mass} -~\ref{np125_contributions_angle}). The different yields therefore illustrate the different acceptance correction factors. The same features as described in Sec.~\ref{InvMass} and Sec.~\ref{AngDistr} for the results inside acceptance can be observed for the results extrapolated in $4\pi$. Indeed the OPER \cite{oper2} and modified Valencia \cite{wasa-3_nnpi+pi+} models give the best overall description of the shape of both distributions.
The OPER \cite{oper2} and in a lesser extent the modified Valencia \cite{wasa-3_nnpi+pi+} models give  also good results for the $\pi^{+}\pi^{-}$ invariant mass and opening angle distributions measured in the $np \to np\pi^{+}\pi^{-}$ at 1. and 1.5 GeV. 
The results obtained at the three energies allow to follow the evolution of the mechanisms, characterized in particular by the important role of non-resonant contributions at 1 GeV and the dominance of double Delta excitation at 1.5 GeV. As for the $np \to  np\pi^{0}\pi^{0}$ reaction at 1.13 GeV \cite{wasa2015-pi0pi0}, where a dominance of the $d*$ contribution is observed, the shape of the $\pi\pi$ invariant mass is similar to our measurement, which confirms the rather low sensitivity of this observable to the s-channel resonant contribution, as stressed by the authors.

\begin{figure}[t]
	\centering
		\includegraphics[width=90mm]{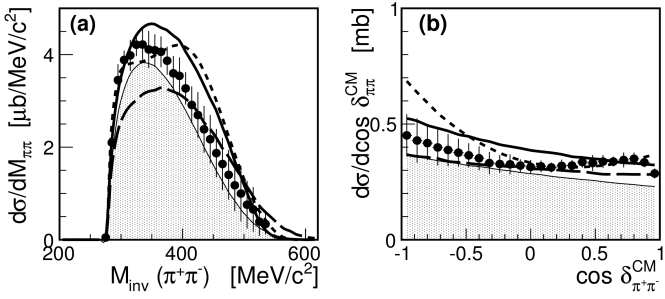} 
	\caption{The acceptance corrected distributions of the $\pi^{+}\pi^{-}$ invariant mass (a) and opening angle of $\pi^{+}\pi^{-}$ in the $np$ rest frame (b) for the $np \to np\pi^+\pi^-$ reaction at 1.25~GeV. The experimental data are shown by solid symbols. The theoretical predictions in $4\pi$ region from OPER \cite{oper2}, Cao \cite{china} and modified Valencia models \cite{wasa-3_nnpi+pi+} are given by the solid,  dashed and  long-dashed curves, respectively. The shaded areas show the phase-space distributions.}
	\label{fig:acc_corr_distr}
\end{figure}

\section{Conclusion}
\label{conclusion}

We have presented an analysis of high-statistics differential distributions obtained in the \pnpnpippim\ reaction, using a deuterium beam at 1.25 GeV/nucleon and the HADES experimental set-up at GSI.  In this channel, in addition to t-channel processes, mainly due to the Roper and double \Del (1232) excitations, the contribution of a s-channel process, with an intermediate dibaryon, as observed in the reaction $pn \to d$\pip\pim\ might be present. In this paper,  we focused on the compatibility of our data with models including  t-channel processes only. The experimental distributions have therefore been compared with the predictions of two Lagrangian models, the Cao model \cite{china} and the "modified Valencia" model \cite{wasa-3_nnpi+pi+} and of the OPER model \cite{oper2}, based on one-pion and one-baryon exchanges. This analysis demonstrates the high sensitivity of the differential distributions to the different components of the two-pion production mechanism. 
The modified Valencia model gives a much better description of the shapes of the distributions than the Cao model. This good result  supports the changes introduced by the WASA collaboration to describe the total and differential cross sections in $pp \to pp$\piz\piz\ and $pp \to$ nn\pip\pip\ in the same energy range, especially the  reduction of the $N(1440) \to$\Del (1232) $\pi$ contribution. The double \Del\ contribution seems to be rather well described by the model, which is important since this process is expected to contribute to dilepton production in the $pn$ reaction \cite{Bashkanov_dileptons13}. However, some discrepancies are observed in our channel, like a missing strength at small \pip\pim\ invariant masses  and forward  $\theta^{CM}_{p\pi^+}$ angles in the model. These discrepancies could be probably reduced by further adjusting the \Del (1600) and Roper contributions. However, such changes should also be validated by an analysis of two-pion production in other isospin channels. The OPER model gives also rather good results,  except for the polar angle of the $p$\pip\ system in the $np$ rest frame, with a lower relative contribution of the double \Del(1232) contribution, and a significant contribution of  one-baryon exchange graphs. Both models have been used to calculate acceptance corrections, and obtain an estimate  of the total cross section of the \pnpnpippim\ reaction at 1.25 GeV beam energy. This measurement is important due to the scarce existing measurement in the relevant energy range. The found  rather low value of the cross section is not inconsistent with a resonant structure at low energies, as expected in presence of the dibaryon resonance, with mass M$\sim$ 2.38 GeV and width $\Gamma\sim$ 70 MeV reported by the WASA collaboration. However, our measurement is hardly compatible with the measurement performed at KEK at an incident neutron  energy lower by 70 MeV. The modified Valencia model, which has been now validated for different channels, underestimates the total cross section in our measurement by only 30$\%$. Under the resonance peak, the underestimation of the data is much larger and is also compatible with the resonant hypothesis. However, the present situation, both from  experimental and theoretical aspects, is not clear enough to draw conclusions on the existence of the dibaryon resonance.
On the experimental side, useful constraints  in the present experiment could be obtained from the extraction of cross section at different neutron energies, the on-going analysis of the $pp \to  pp$\pip\pim\ and $pn \to d$\pip\pim\ channels with HADES will allow for specific tests of the modified Valencia model and the dibaryon resonance contribution, respectively. The experimental situation should therefore further become clearer in a near future. On the theory side, we 
would like to call for the development of a full model, including in a consistent way the t-channel processes, based on the modified Valencia model and the s-channel processes including the dibaryon with above quoted parameters, which could provide a solid framework for the interpretation of the two-pion production data.

\vskip 5mm
We acknowledge valuable discussions with Dr. T.~Skorodko and are particularly indebted to her and
Dr. Xu Cao for the provided calculations. 
The collaboration gratefully acknowledges the support by 
BMBF grants 06TM970I, 06GI146I, 06FY9100I, and 05P12CRGHE (Germany),
by GSI (TM-FR1, GI/ME3, OF/STR), by Helmholtz Alliance
HA216/EMMI, by grants MSMT LC07050, LA316 and GA ASCR
IAA100480803 (Czech Republic), by grant NCN 2013/10/M/ST2/00042
(Poland), by INFN (Italy), by CNRS/IN2P3 (France), by grants MCYT
FPA2000-2041-C02-02 and XUGA PGID T02PXIC20605PN (Spain),
by grant UCY-10.3.11.12 (Cyprus), by INTAS grant 03-51-3208 and
by EU contract RII3-CT-2004-506078.



%% References


%\bibliographystyle{elsarticle-num}



\end{document}